\documentclass[twocolumn, aps]{revtex4-2}
\usepackage{graphicx}
\usepackage{amsmath}
\usepackage{graphics}

\usepackage{amsmath}
\usepackage{amsfonts}
\usepackage{amssymb}
\usepackage{xcolor}
\usepackage{subcaption}
\DeclareUnicodeCharacter{2212}{-}

\usepackage[normalem]{ulem}

\begin{document}

\title{A physically realizable molecular motor driven by the         Landauer blowtorch effect}

\author{Riley J. Preston}
\affiliation{Institute of Physics, University of Freiburg, Hermann-Herder-Strasse 3, 79104 Freiburg, Germany }

\author{ Daniel S. Kosov}
\email{Authors to whom correspondence should be addressed: daniel.kosov@jcu.edu.au}
\affiliation{College of Science and Engineering, James Cook University, Townsville, QLD, 4811, Australia }

\begin{abstract}
We propose a model for a molecular motor in a molecular electronic junction driven by a natural manifestation of Landauer's blowtorch effect. The effect emerges via the interplay of the electronic {   friction} and diffusion coefficients, each calculated quantum mechanically using nonequilibrium Green's functions, within a semi-classical Langevin description of the rotational dynamics. The motor functionality is analysed through numerical simulations where the rotations exhibit a directional preference according to the intrinsic geometry of the molecular configuration. The proposed mechanism for motor function is expected to be ubiquitous for a range of molecular geometries beyond the one examined here.
\end{abstract}

\maketitle

\newpage
\section{INTRODUCTION}

Experimental demonstrations of molecular motors have used a range of external energy sources such as light \cite{wilcken18,kistemaker15,balzani06,vandelden05,koumura99}, chemical reactions \cite{kinosita00,leigh03,juluri09}, thermal gradients \cite{barreiro08}, or applied electric currents \cite{stolz20,ren20,zhang19,eisenhut18,mishra15,perera13,tierney11,kudernac11}, the latter being of particular interest due to its conceptual compatibility with nanoelectronics. With this in mind, this work considers a molecular rotor subject to an applied electric current as supplied by a pair of conducting electrodes. The molecular rotor serves as the main conducting element in a molecular electronic junction, capable of producing mechanical work.

There already exists a wealth of theoretical literature describing such systems whose motor functionalities arise from a range of physical phenomena, including, but not limited to, quantum tunneling \cite{stolz20} and excitation-relaxation \cite{ribetto22,echeverria14} processes in asymmetric ratchet potentials, non-Markovian behaviour of the current-induced forces leading to a bias in the directionality \cite{nie09}, {   and non-conservative forces \cite{calvo17,dundas09}}. However, previous studies have overlooked the possible functionality which can arise due to the inhomogeneous dissipative-excitational current-induced forces present in such systems.

In this paper, we consider a model in which a molecular rotor is driven by the current-induced forces imparted by electrons tunneling through 
  it. These forces provide the required  energy to the rotational degree of freedom in order to overcome the potential barrier for rotation. In parallel with previous work \cite{ribetto22,calvo17,echeverria14}, we model the rotational degree of freedom classically according to a Keldysh-Langevin approach where its time-evolution is governed by three components; an adiabatic force which sets the shape of the ratchet potential, as well as a dissipative frictional force and a stochastic force, the balance of which yields the steady-state temperature of the classical {        rotator}. Each of these forces, which arise due to the interaction with the quantum nonequilibrium electronic environment, is calculated self-consistently via nonequilibrium Green's functions. 

The directionality of our rotation is a result of - to our knowledge - a hitherto unexplored contribution for motors in molecular junctions, that being a consequence of Landauer's blowtorch effect \cite{Landauer1975,Landauer1993} in the ratchet potential, which emerges via the interplay of the coordinate-dependent diffusion and viscosity coefficients. This phenomena is well-understood in the context of chemical reaction rates \cite{preston21}, whereas here the scope is extended to the study of ratchets. This is in contrast to previous research where the directional rotation comes as a result of the non-conservativity of the adiabatic force \cite{ribetto22,calvo17} - a phenomena which is also easily accessible with our model via an appropriate choice of Hamiltonian, but is not the aim of this study. {        Driving the motor by the blowtorch effect} is of particular interest since the dissipative and stochastic forces generally act to degrade the device performance rather than enforce it \cite{calvo17}. We note that while the effect of inhomogeneous viscosity and diffusion coefficients in ratchets has been explored on a mathematical level \cite{buttiker87,luchsinger00}, here we propose a physically realizable molecular electronic junction in which the effect emerges naturally. This effect does not require an explicit time-dependence of the Hamiltonian as it arises due to the molecular geometry. Additionally, the directional rotation does not require an asymmetric ratchet potential, although such asymmetric potentials can arise from our calculations via the adiabatic force, 
{        further reinforcing motor performance}. We note {        that the function of our motor is reliant on the rotational dynamics being sufficiently damped; a regime which is generally  fulfilled in molecular electronic junctions since the conducting molecule is usually embedded into an insulating solvent or it is a part of a molecular monolayer.}

{   It has been shown theoretically that a non-zero charge current can be pumped through a quantum system in equilibrium via the periodic variation of two independent parameters \cite{brouwer98}. A particularly relevant example is described in Ref.\cite{yadalam16}, where the coupling of a quantum system to the left and right electrodes each assumes a  periodic time-dependence. Since in our model the coupling to each lead is implicitly time-dependent through the evolution of the nuclear geometry, this would be equivalent to a constant, manual rotation of our molecular motor with constant angular velocity. We use our model to investigate the converse effect to equilibrium charge shuttling, in which an applied charge current via the nonequilibrium electrodes \emph{produces} a time-dependent variation of two independent parameters (the coupling to the left and right electrodes) which emerges via the directed rotation of the molecular geometry. Thus, this is an example of an adiabatic quantum motor. We do, however, find that the operational parameter regimes of our molecular motor differ from that of models of equilibrium charge shuttling, which we further discuss in the results section.} Our choice of Hamiltonian also mirrors an example demonstrated in Ref. \cite{cizek11}, where the rotation is instead considered from a quantum perspective.

\section{MODEL}
\label{Theory}

\begin{figure}
\includegraphics[width=0.5\textwidth]{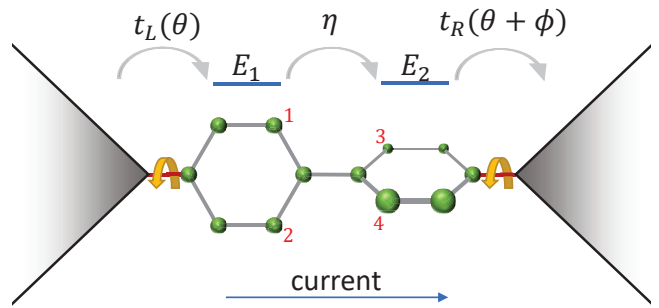}
\caption{Schematic of the model system. A biphenyl based molecule (green) connects two graphene electrodes. The molecule represents a rigid  rotator. The         dihedral  angle between the two phenyl rings, which is critical for motor functionality, can be adjusted by the inclusion of appropriate side groups to atoms 1,2,3, and 4. An applied current induces a directional rotation about the red bonds when the dihedral angle between the phenyl rings is non-zero.}
\label{system_diagram}
\end{figure}

A visualisation of our proposed molecular junction configuration is shown in Fig. \ref{system_diagram}. We have two planar electrodes bridged by a biphenyl based molecule. The phenyl rings are prepared such that they are displaced by a         dihedral angle $\phi$ from each other - this angle is a constant for a given simulation. The motor effect arises through the angle $\theta$, which represents the uniform rotation of the entire molecular bridge        as a rigid body. To produce an observable directionality to the rotations, the vibrations must be adequately damped. We find that the electronically calculated forces are generally insufficient to achieve this regime and so we additionally include an external equilibrium environment; for example, a solvent or        molecular monolayer surrounding the junction, which acts to further dampen the classical vibrations. We emphasize that the proposed geometry is merely a physically reasonable suggestion. The proposed motor effect should be ubiquitous in molecular junction geometries provided that there is an asymmetry in the Hamiltonian, in our case arising from the        dihedral angle $\phi$.

The system is described by a generic tunneling Hamiltonian as per
\begin{equation}
\hat{H}(t) = \hat{H}_{M} + \hat{H}_{L} + \hat{H}_{R} + \hat{H}_{LM}(\theta(t)) + \hat{H}_{MR}(\theta(t)) + H_{\text{cl}}(t).
\label{hamiltonian}
\end{equation}
The total system Hamiltonian is partitioned into the following components; the molecular Hamiltonian $\hat{H}_{M}$ for the molecular bridge, the left and right {         electrodes }Hamiltonians $\hat{H}_{L}$ and $\hat{H}_{R}$, the          electrodes-molecule coupling Hamiltonians $\hat{H}_{LM}(\theta(t))$ and $\hat{H}_{MR}(\theta(t))$ which describe the coupling between the electronic states on the rotor and the left and right {         electrodes}, respectively, and the classical Hamiltonian $H_{\text{cl}}(t)$ which describes the time-evolving molecular geometry. Note that $\hat{H}_{LM}(\theta(t))$ and $\hat{H}_{MR}(\theta(t))$ depend on time implicitly via the classical, rotational degree of freedom $\theta$.

The molecular Hamiltonian consists of two conducting electronic levels, each localised on one of the {        phenyl} rings. It then takes the form
\begin{equation}
\hat{H}_{M} = E_{1} \hat{d}^{\dag}_{1} \hat{d}_{1} + E_{2} \hat{d}^{\dag}_{2} \hat{d}_{2} + v (\hat{d}^{\dag}_{1} \hat{d}_{2} + \hat{d}^{\dag}_{2} \hat{d}_{1}).
\label{molecularhamiltonian}
\end{equation}
$E_1$ and $E_2$ are the energies of the first and second electronic levels, respectively, while $v$ is the hopping amplitude.

The {         electrodes }are described as non-interacting fermionic baths and the Hamiltonian is taken in the standard form,
\begin{equation}
\hat{H}_{L} + \hat{H}_{R}  = \sum_{k \alpha} \epsilon_{k \alpha} \hat{d}^{\dagger}_{k \alpha} \hat{d}_{k \alpha},
\end{equation}
where we use a subscript $k\alpha$ to denote an operator acting on state $k$ in the $\alpha$ electrode which has energy $\epsilon_{k \alpha}$.

The {        molecule-electrode coupling}, $\hat{H}_{LM}$ and $\hat{H}_{MR}$, are defined according to
\begin{align}
\hat{H}_{LM} &= \sum_{k\in L} \Big( t_{k 1}(\theta (t) )  \hat{d}^{\dagger}_{k } \hat{d}_{1} + \text{h.c.} \Big), \\
\hat{H}_{RM} &= \sum_{k\in R} \Big( t_{k 2}(\theta (t) + \phi )  \hat{d}^{\dagger}_{k } \hat{d}_{2} + \text{h.c.} \Big).
\end{align}
The matrix elements $t_{k i}$ (and their conjugates) describe the tunneling amplitudes between {        electrode} states $k$ and 
the {        molecular bridge states}  $i$, where state $1$ is only coupled to the left electrode and state $2$ is only coupled to the right {        electrode}. Note that $t_{ki}$ depends explicitly on the classical rotational coordinate $\theta$. We choose to express $t_{k\alpha,i} (\theta) = t_{k\alpha,i} s_\alpha (\theta)$, where the classical dependence emerges through $s_\alpha (\theta)$, which takes the following forms for the left and right {         electrodes}: 
\begin{align}
s_L &= 1 + \frac{A}{2}\left( \cos(2\theta)-1\right),\\
s_R &= 1 + \frac{A}{2}\left(\cos(2(\theta + \phi))-1\right).
\end{align}
With this dependence, the coupling amplitude is maximised when a phenyl ring is coplanar with its corresponding electrode and minimised when the phenyl ring is orthogonal to the electrode with a magnitude of $1-A$ times the maximum value.   This dependence of the tunneling amplitudes on the rotational angle can be realized physically using  graphene electrodes, where the rotation of the molecular bridge out of the electrode plane lowers $\pi$-conjugation, reducing the corresponding tunneling amplitude.

Finally, the classical Hamiltonian is given by a rigid rotator expression,
\begin{equation}
H_{\text{cl}}(t) = \frac{L^2}{2I} + U_\text{cl}(\theta),
\end{equation}
where $L$ is the angular momentum of the molecular geometry, $I$ is the moment of inertia and $U_\text{cl}(\theta)$ is the classical potential for the rotation. In our calculations we set $U_\text{cl}(\theta) = 0$, such that the {        rotational} potential results entirely from the interaction with the electronic environment, calculated quantum mechanically. In any case, the inclusion of a non-zero classical potential will not have a qualitative difference on the observed motor effect. 

\section{Current-induced torque and "blowtorch" temperature}

The operator for the torque acting on the classical rotational coordinate due to the quantum, electronic environment is given by
\begin{equation}
\hat{\tau} =-\partial_\theta \hat{H}(t)
=-\sum_{k\alpha, i}\left[\partial_\theta t_{k\alpha i}(\theta) \hat{d}_{k\alpha}^{\dagger}\hat{d}_{i} +h.c. \right],
\label{heisenberg}
\end{equation}
where $\partial_{\theta}$ is the partial derivative with respect to $\theta$. 
The summation in the above runs over both         electrodes, $\alpha \in \{L,R\}$, and both {        molecular} electronic states, $i \in \{1,2\}$. The torque operator is then expressed in terms of a mean term and a deviation from the mean,
\begin{equation}
\hat{\tau} = \langle \hat{\tau} \rangle + \delta \hat\tau,
\label{partition}
\end{equation}
where each can be quantified in terms of nonequilibrium Green's functions. As is covered in detail in the appendix, a time-scale separation between the slow classical rotation of the rotor and the fast electron tunneling allows for a perturbative expansion of the mean torque in terms of the small parameter {       - the derivative with respect to central time in the molecular bridge Green's functions. The perturbative expansion is
\begin{equation}
\langle \hat{\tau} \rangle = \tau_{(0)}(\theta) + \tau_{(1)}(\theta,\dot\theta) + ...,
\end{equation} 
where $\tau_{(n)}$ is of $n^\text{th}$ order in {        the central time derivatives}. We truncate the expansion after the first order. We calculate a conservative potential according to 
\begin{equation}
U = -\int^\theta_{\theta_0} d\theta '\tau_{(0)}(\theta'),
\label{pot}
\end{equation}
where the choice of {        $\theta_0$} is arbitrary. Equation (\ref{pot}) entirely defines the ratchet potential for our rotational coordinate due to the electronic environment.
Finally, the torque operator is then mapped onto a classical torque such that we obtain a classical equation of motion for the rotational coordinate. It takes the form of a Langevin equation,
\begin{equation}
I\ddot{\theta} = \tau_{(0)}(\theta) - (\xi_{\text{solv}} + \xi (\theta)) \dot{\theta} + \delta \tau(t),
\label{lang_eqn}
\end{equation}
where $\xi(\theta)$, calculated via $\tau_{(1)}$, is the electronic {   friction} coefficient while $\xi_{\text{solv}}$ is the friction due to the interaction with an external solvent. $\delta\tau(t)$ is a classical stochastic force quantified according to a diffusion coefficient, $D_\text{tot} = D(\theta) + D_\text{solv}$, where the electronic part is defined according to 
\begin{equation}
\langle \delta \tau(t) \delta \tau(t') \rangle = D(\theta) \delta (t-t').
\end{equation}
Each of the electronic forces, $\tau_{(0)}(\theta)$, $\xi(\theta)$, and $D(\theta)$, are calculated quantum mechanically via nonequilibrium Green's functions while the forces due to interaction with the external solvent, $\xi_\text{solv}$ and $D_\text{solv}$, are input parameters to the model which allow us to artificially increase the damping of the dynamics. {   We have applied the white-noise approximation in calculating the electronic part of the diffusion coefficient which is justified due to the clear separation of time-scales between the electronic and classical dynamics \cite{preston22}. The same cannot be said for the external damping, whose dynamics may occur on similar time-scales to the classical rotations. However, the operation of our motor is governed chiefly by the behaviour of the electronic component and as such, we predict that a more accurate approach to the modelling of the external solvent is not important for the observed motor effect.} In analogy with the fluctuation-dissipation theorem, we can define an effective "blowtorch" temperature for the classical rotation  according to \cite{zwanzig-book,preston2020,preston21,prestonAC2020}
\begin{equation}
k_B T_\text{eff}(\theta) = \frac{D(\theta) + D_\text{solv}}{2(\xi(\theta) + \xi_\text{solv})}.
\end{equation}
      Expressions for the diffusion coefficient, viscosity and average torque in terms of nonequilibrium Green's functions along with relevant derivations are given in the appendix.

\section{RESULTS}
\label{Results}

We now present results for our model system. Results are acquired via computational simulations of the Langevin dynamics produced by our model according to (\ref{lang_eqn}). From long Langevin trajectories in time, we calculate the average rotation rate for the classical rotational degree of freedom for a chosen set of parameters. The common parameters for all calculations, unless otherwise specified, are as follows. The {         electrode }temperatures are set such that $k_B T_\alpha \approx 2.72 \times 10^{-2} \text{eV}$, and the solvent is in thermal equilibrium with the {         electrodes }with a corresponding viscosity coefficient of {   $\xi_{\text{solv}}= 5 $~a.u.}. The moment of inertia of the classical rotational coordinate is approximated according to two phenyl rings as {   $I=4.5\times10^5$ a.u.}. 
We use the wide-band approximation, 
and express the level broadening as
\begin{equation}
\Gamma_{\alpha} =\Gamma_\alpha^\text{max} s_\alpha^2 (\theta),
\end{equation}
where the maximum level broadenings $\Gamma_\alpha^\text{max}$ are input parameters in our calculations.
We take the maximum level broadenings due to the left and right          electrodes, respectively, as 
$\Gamma^\text{max}_L \approx 0.272 \text{eV}$ and $\Gamma^\text{max}_R = \Gamma^\text{max}_L/2$. For the molecular Hamiltonian, we take $E_1 = E_2 = 0$ while the hopping amplitude is given by {   $v = 1.25 \text{eV}$}. {   We apply the voltage, $\mathcal{V}$,} symmetrically in all cases such that $\mu_L = -\mu_R$. Finally, we take $A=0.95$, such that the Hamiltonian coupling element when the phenyl ring is perpendicular to the electrode is $5\%$ of the corresponding coplanar value.

\begin{figure}
\includegraphics[width=0.5\textwidth]{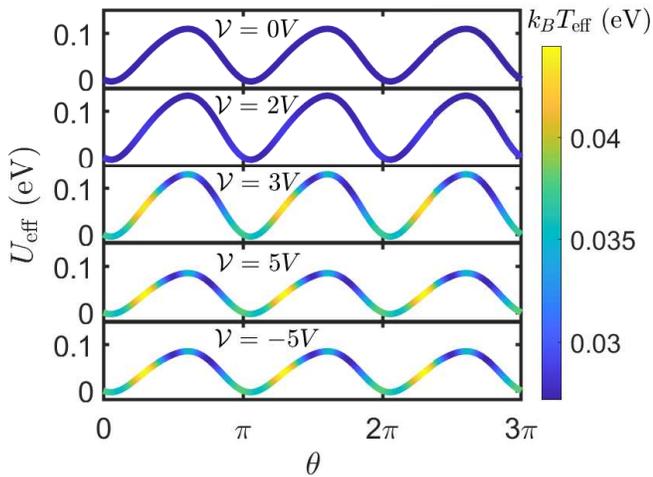}
\caption{$U$ calculated according to (\ref{pot}) with $k_B T_{\text{eff}}$ overlaid on top for different voltages.     
The nonhomogeneous temperature with local hot spots is the manifestation of the Landauer blowtorch effect.
    Dihedral angle between phenyl rings: $\phi = -\pi/4$.}
\label{U_V_fig}
\end{figure}

\begin{figure}
\includegraphics[width=0.5\textwidth]{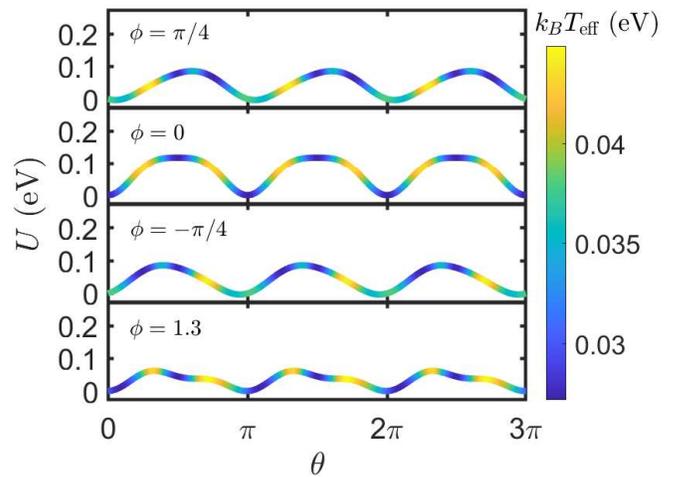}
\caption{$U$ calculated according to (\ref{pot}) with $k_B T_{\text{eff}}$ overlaid on top for different {        dihedral angle between phenyl rings} $\phi$. Voltage $\mathcal{V} = 5V$.}
\label{U_phi_fig}
\end{figure}

{   In Fig. \ref{U_V_fig}, we observe the periodic ratchet potentials generated for a range of voltages along with the corresponding inhomogeneous effective temperatures overlaid on top. At equilibrium, the rotational coordinate is in thermal equilibrium with the electrodes and solvent. Of principal importance are the energies of the molecular orbitals which are $\pm 1.25eV$ for our parameters, which are off-resonant when $\mathcal{V}<2.5V$. Increasing the voltage in the off-resonant regime - exemplified by the $\mathcal{V}=2V$ case - increases the height of the energy barrier for rotation while the temperature of the rotational coordinate differs only slightly from equilibrium. Conversely, in the resonant regime when $\mathcal{V} > 2.5V$, the inhomogeneous temperature as a function of $\theta$ yields clear periodic hot-spots which we refer to as the blowtorch. Further increasing the voltage magnifies these hotspots while decreasing the energy barrier for rotation.} The value of $\phi=-\pi/4$ was chosen specifically here to illustrate a situation in which a periodic blowtorch increases the probability for the forwards rotation (increasing $\theta$). This is because the effects of the potential gradient are nullified in the region where the blowtorch is applied, resulting in an effective decrease to the barrier for rotation in the forwards direction \cite{preston21}. We also observe numerically that our Langevin coefficients are independent of the sign of the voltage. Thus, the rotational direction must also be independent of the sign of the voltage. In other words, our mechanism for the rotation of the molecular structure is independent of the direction of electron tunneling through the junction. If we take this to be true, this then justifies our decision to have $\Gamma^\text{max}_L \ne \Gamma^\text{max}_R$, since otherwise the symmetry of the system would prevent any non-zero average rotation.

\begin{figure}
\includegraphics[width=0.5\textwidth]{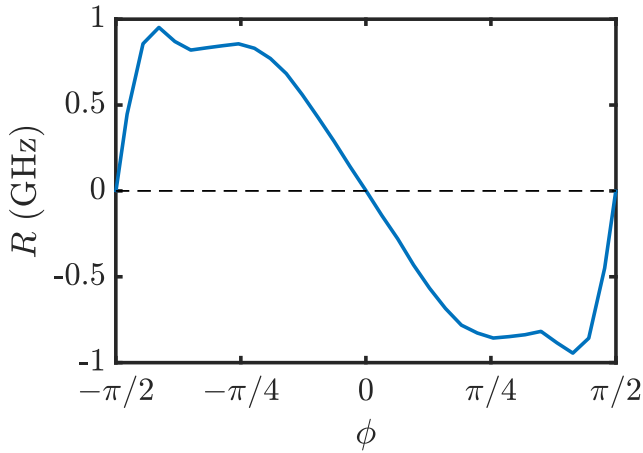}
\caption{The rotation rate, $R$, as a function of {        dihedral angle between phenyl rings} $\phi$. Each $R$ point is calculated via averaging over a trajectory with a length of $\approx 1.6\times 10^6 \; \mathrm{ns}$. Voltage $\mathcal{V}=5V$.}
\label{phase_fig}
\end{figure}

We additionally observe the dependence on $\phi$ in Fig. \ref{U_phi_fig}. When $\phi=0$ and the two {        phenyl}  rings are coplanar, the ratchet potential and corresponding effective temperature distribution are symmetric, ruling out any possible rotation as is to be expected. Upon comparing $\phi=\pi/4$ with $\phi = -\pi/4$, corresponding to opposite chiralities of the molecular bridge, we observe the dependence on $\theta$ to be flipped such that we should observe equal and opposite average rotation rates - a result which we observe directly in Fig. \ref{phase_fig}. The case of $\phi = 1.3$ was chosen to highlight the possibility of deformation to the potential which can have a significant effect on the rotation rate.

We now turn to numerical simulations of the dynamics. In Fig. \ref{phase_fig}, we observe the average rotation rate, $R$, over a trajectory as a function of $\phi$. The rotations go to zero when $\phi = 0$ and  $\pm\frac{\pi}{2}$, as is to be expected from symmetry arguments. We find that $R(-\phi) = -R(\phi)$ as expected from the previous discussion. Short example trajectories of the rotational coordinate as a function of time are plotted for different values of $\phi$ in Fig. \ref{traj_fig}, for the readers intuition.

In Ref. \cite{yadalam16}, equilibrium charge shuttling was shown to be maximised when $\phi = \pm \pi/4$; a result which we can readily reproduce by applying a manual rotation to $\theta$ such that it increases or decreases linearly with time. We find here that the rotation rate due to an applied voltage follows a similar trend, reaching a minimum/maximum at $\phi=\pm\pi/4$. However, we observe a deviation from this behaviour around $\phi=\pm 1.3$ due to the rapid {        current-induced} deformation of the ratchet potential. We also note that equilibrium charge shuttling can even be observed even when $\Gamma^\text{max}_L = \Gamma^\text{max}_R$; a regime in which we do not observe a net rotation by applying a voltage since our mechanism for rotation is independent of the direction of the current. {   In contrast, models for equilibrium charge pumping show that the produced current is \emph{reversed} upon reversing the rotation of the molecular configuration \cite{yadalam16}. We find that the direction of rotation in our model is determined by the choice of $\phi$ as well as the choices of $\Gamma_L^\text{max}$ and $\Gamma_R^\text{max}$. We have arbitrarily chosen $\Gamma_L^\text{max}>\Gamma_R^\text{max}$ to produce the displayed results. If we instead choose $\Gamma_R^\text{max}>\Gamma_L^\text{max}$, the observed rotational directions are reversed - a result we have observed numerically but not shown here.}

\begin{figure}
\includegraphics[width=0.5\textwidth]{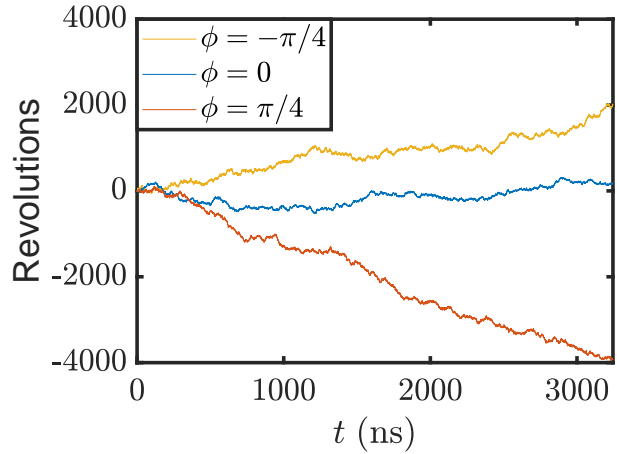}
\caption{Short time trajectories of the         rotational angle  $\theta$ (expressed here in terms of the number of revolutions) for different values of        the dihedral angle $\phi$.    Voltage $\mathcal{V} = 5V$.}
\label{traj_fig}
\end{figure}

{   Fig. \ref{voltage_fig} demonstrates the voltage dependence of the rotational rate. We observe negligible rotation in the off-resonant regime when $\mathcal{V}<2.5V$. In the resonant regime, the average rotation rate increases approximately linearly due to the increasing magnitude of the applied blowtorch with increasing voltage along with the lowering of the energy barrier required for rotation.} For even higher voltages, we expect that the rotation rate would begin decreasing back towards zero since the large effective temperatures will overwhelm the potential entirely, removing any directional preference. This, however, would occur beyond the realms of physically achievable voltages for our model.

\begin{figure}
\includegraphics[width=0.5\textwidth]{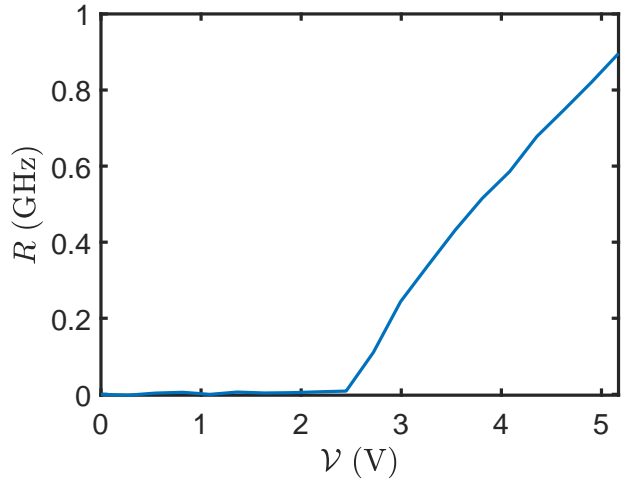}
\caption{The rotation rate, $R$, as a function of {        voltage} $\mathcal{V}$. Each $R$ point is calculated via averaging over a trajectory with a length of $\approx 1.6\times 10^6 \; \mathrm{ns}$. {        Dihedral angle} $\phi = -\pi/4$.}
\label{voltage_fig}
\end{figure}
\begin{figure}
\includegraphics[width=0.5\textwidth]{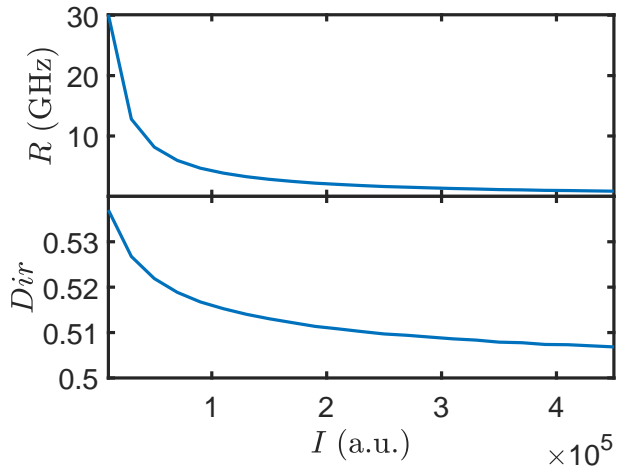}
\caption{The rotation rate $R$, and directionality, $Dir$, as a function of the moment of inertia of the classical rotational coordinate. The final point on each plot corresponds to our usual choice of $I$ for two phenyl rings. The trajectory length was chosen for each value of $I$ to ensure convergence of the results.          Dihedral angle $\phi = -\pi/4$,        voltage $\mathcal{V} = 5V$.}
\label{inertia_fig}
\end{figure}

The function of our molecular motor requires sufficient damping - a regime we achieve via the inclusion of an external solvent to the system. In Fig. \ref{inertia_fig}, we observe the dependence of the rotation rate on the moment of inertia of the molecular configuration, where $I \approx 1.15 \times 10^{-45} \text{kgm}^2$ is the physically reasonable value corresponding to our chosen molecular configuration. In the overdamped case where $I$ is unrealistically small, the rotation rate is orders of magnitude larger than for realistic values for $I$. The rotation rate asymptotically decreases towards zero with increasing moment of inertia, where in the underdamped case, the preference of a given direction will become vanishingly small. As an additional insight, we define the directionality according to 
\begin{equation}
Dir = \frac{n_{\text{forw}}}{n_{\text{forw}} + n_{\text{back}}},
\end{equation}
where $n_{\text{forw}}$ and $n_{\text{back}}$ are the number of forward and backward rotations over the full length of the trajectory. $Dir = 1$ would correspond to a trajectory in which the molecular motor rotates unidirectionally forwards. For the physically realistic value of $I$, $50.68\%$ of all rotations are forwards. This directionality is far smaller than what has been demonstrated for motors governed chiefly by quantum effects \cite{stolz20}.

\section{CONCLUSIONS}

In this paper, we have proposed an experimentally realizable model for a molecular motor in a molecular electronic junction whose operation is governed by Landauer's blowtorch effect. This contrasts with other theoretical models for molecular motors which generally disregard the inhomogeneous temperature of the electronic environment induced by the nonequilibrium          electrodes. We have demonstrated that directional rotations can be produced entirely as a result of the behaviour of the viscosity and diffusion coefficients -         these are exerted by tunneling quantum  electrons on the classical rotator and  calculated exactly via nonequilibrium Green's functions - while the         rotational potential is periodic and subsequently introduces no intrinsic directionality of its own. This effect is, however, limited to regimes where the        rotations are sufficiently damped and we anticipate that the small electronic {   friction} alone will not be enough to produce a non-negligible rotational preference, hence our choice to additionally include an external solvent which increases the dampening of the rotation.

\begin{center}
{\bf DATA AVAILABILITY}
\end{center}

The data that supports the findings of this study are available within the article.

\clearpage

\appendix
\begin{widetext}
\section{Torque, Rotational Viscosity and Diffusion Coefficient in Terms of Nonequlibrium Green's functions}
We use the standard definitions for the lesser $G_{ij}^<(t,t') $, greater $G_{ij}^>(t,t')$, retarded $G_{ij}^R(t,t')$ and advanced $G_{ij}^A(t,t')$ components of the electronic Green's functions in our derivations. 
Expressing the torque operator in the Heisenberg picture, we compute the average torque as
\begin{equation}
\langle \hat{\tau} \rangle = i\sum_{k\alpha i}\left[\partial_\theta t_{k\alpha i}(\theta)G_{ik\alpha}^{<}(t,t)+\partial_\theta t_{ik\alpha}(\theta)G_{k\alpha i}^{<}(t,t)\right].
\label{mean_torque_exact}
\end{equation}
This torque is computed for the exact, nonadiabatic Green's functions.

We now perform a perturbative expansion of the mean torque given in (\ref{mean_torque_exact}). It is a mathematical convenience to perform this expansion under a Wigner transformation of the time since it allows for the easy recognition of different time-scales within the system. The Wigner time coordinates are defined according to
\begin{equation}
T = \frac{t + t'}{2},
\;\;\;\;\;\;\;
\tau = t - t',
\end{equation}
where $T$ is the central time, associated with the long time-scales of classical vibration and $\tau$ is the relative time, related to electronic tunneling. Thus, in our theory the small parameter naturally emerges via derivatives with respect to $T$. We introduce an auxilliary two-time function,
\begin{equation}
\mathcal{T} (t,t') = i\sum_{k\alpha i}\left[\partial_\theta t_{k\alpha i}(\theta(t'))G_{ik\alpha}^{<}(t,t')+\partial_\theta t_{ik\alpha}(\theta(t))G_{k\alpha i}^{<}(t,t')\right],
\end{equation}
where $\mathcal{T} (t,t) = \langle \hat{\tau}  (t) \rangle$. Next, the Green's functions spanning both the electrode and molecular space can be decomposed via the Dyson equation
\begin{equation}
G^{<}_{k\alpha i}(t,t')=\int_{-\infty}^{\infty}dt_1\sum_{j}\left[g_{k\alpha }^{<}(t,t_1)t_{k\alpha j}(t_1)G^{A}_{j i}(t_1,t')\right.
+\left.g_{k\alpha }^{R}(t,t_1)t_{k\alpha j}(t_1)G^{<}_{j i}(t_1,t')\right],
\label{dyson_expansion}
\end{equation}
where $g_{k\alpha }(t,t_1)$ is the free Green's function for electrode $\alpha$.
 The resultant equation for $\mathcal{T} (t,t')$ is
\begin{equation}
\mathcal{T} (t,t')  =i\sum_{ij}\int_{-\infty}^{\infty}dt_{1}\Big[G_{ij}^{<}(t,t_{1})\Phi_{ji}^{A}(t_{1},t')+G_{ij}^{R}(t,t_{1})\Phi_{ji}^{<}(t_{1},t')
  +\Psi_{ij}^{<}(t,t_{1})G_{ji}^{A}(t_{1},t')+\Psi_{ij}^{R}(t,t_{1})G_{ji}^{<}(t_{1},t')\Big].
\label{torque_aux}
\end{equation}
Here we have introduced the self-energy-like terms, $\Psi$ and $\Phi$, which contain any information about the coupling to the electrodes. These are defined as ($c=<,>, R, A$)
\begin{equation}
\Psi^{c}_{ij}(t,t') = \sum_{k\alpha} \partial_\theta t_{ik\alpha}(\theta(t))g_{k\alpha }^{c} (t,t') t_{k\alpha j}(\theta(t')),
\end{equation}
\begin{equation}
\Phi^{c}_{ij}(t,t') = \sum_{k\alpha} t_{ik\alpha}(\theta(t)) g_{k\alpha}^{c} (t,t') \partial_\theta t_{k\alpha j}(\theta(t')).
\end{equation}
Application of the Wigner transform to (\ref{torque_aux}) results in    
\begin{equation}
\int d\tau e^{i\omega\tau}\mathcal{T}(t,t') =\text{Tr}\Big\{ie^{\frac{1}{2i}\lambda(\partial_{T}^{G}\partial_{\omega}^{\Phi}-\partial_{\omega}^{G}\partial_{T}^{\Phi})}\left(\tilde{G}^{<}\tilde{\Phi}^{A}+\tilde{G}^{R}\tilde{\Phi}^{<}\right)+ie^{\frac{1}{2i}\lambda(\partial_{T}^{\Psi}\partial_{\omega}^{G}-\partial_{\omega}^{\Psi}\partial_{T}^{G})}\left(\tilde{\Psi}^{<}\tilde{G}^{A}+\tilde{\Psi}^{R}\tilde{G}^{<}\right)
\Big\},
\label{torque_wig}
\end{equation}
where we use $\tilde{G}$ to denote the Wigner transform of $G$, defined as {   
\begin{equation}
\tilde{G}(T,\omega) = \int d\tau e^{i \omega \tau} G(T,\tau),
\end{equation}}
and the same applies for the self-energy-like terms. Functions in the Wigner space carry dependence on $T$ and $\omega$ which we subdue for brevity.  We now propose the ansatzes,
\begin{equation}
\tilde{G}=\tilde{G}_{(0)} + \lambda\tilde{G}_{(1)} + \lambda^2 \tilde{G}_{(2)}+ ...,
\label{ansatz_G}
\end{equation}
\begin{equation}
\tilde{\Psi}=\tilde{\Psi}_{(0)} + \lambda\tilde{\Psi}_{(1)} + \lambda^2\tilde{\Psi}_{(2)}+ ...,
\label{ansatz_psi}
\end{equation}
\begin{equation}
\tilde{\Phi}=\tilde{\Phi}_{(0)} + \lambda\tilde{\Phi}_{(1)} + \lambda^2\tilde{\Phi}_{(2)}+ ...,
\label{ansatz_phi}
\end{equation}
in which $\tilde{G}_{(n)}$ is of $n^{th}$ order in our small parameter, and the same applies to $\tilde{\Psi}$ and $\tilde{\Phi}$. Terms with $n=0$ correspond to the adiabatic approximation, while the higher order terms go beyond this and account for the dynamical corrections due to molecular rotations. 
We use $\lambda$ in the above as a book-keeping term which makes clear the "smallness" of the term in question. For example, a term proportional to $\lambda$ will be first order in our small parameter, and so on. We let $\lambda=1$ at the end of the derivation. 

We substitute these expansions into (\ref{torque_wig}) and consider each order of $\lambda$ separately. In the adiabatic case, we retain only the $n=0$ terms from (\ref{ansatz_G})-(\ref{ansatz_phi}) while the exponentials in (\ref{torque_wig}) disappear, resulting in
\begin{equation}
\int d\tau e^{i\omega\tau}\mathcal{T}_{(0)}(t,t') =i\text{Tr}\Big\{\tilde{G}_{(0)}^{<}\tilde{\Phi}_{(0)}^{A}+\tilde{G}_{(0)}^{R}\tilde{\Phi}_{(0)}^{<}+\tilde{\Psi}_{(0)}^{<}\tilde{G}_{(0)}^{A}+\tilde{\Psi}_{(0)}^{R}\tilde{G}_{(0)}^{<}
\Big\},
\end{equation}
where we have let $\lambda=1$. We then apply the inverse Wigner transform and let $\tau = 0$ which yields
\begin{equation}
\tau_{(0)} =-\int \frac{d\omega}{\pi}\text{ImTr}\Big\{\tilde{\Psi}_{(0)}^{<}\tilde{G}_{(0)}^{A}+\tilde{\Psi}_{(0)}^{R}\tilde{G}_{(0)}^{<}
\Big\}.
\label{adiabatic_force}
\end{equation}
We use $\text{ImTr}$ to denote the imaginary part of the trace, where we have used the fact that $(X^<)^\dag = -X^<$ and $(X^A)^\dag = X^R$ for an arbitrary term $X$. (\ref{adiabatic_force}) specifies the adiabatic torque. 

We now consider the first-order non-adiabatic correction to the average torque. This is found by retaining the first-order terms in (\ref{torque_wig}), which are linear in $\lambda$. With some work, we find
\begin{multline}
\tau_{(1)}  = -\frac{1}{\pi} \int d\omega\text{ImTr}\left\{\tilde{\Psi}_{(0)}^{R}\tilde{G}_{(1)}^{<}+\tilde{\Psi}_{(1)}^{<}\tilde{G}_{(0)}^{A}\right. 
\left.+\tilde{\Psi}_{(1)}^{R}\tilde{G}_{(0)}^{<}+\tilde{\Psi}_{(0)}^{<}\tilde{G}_{(1)}^{A}\right\} \\
  +\frac{1}{2\pi}\int d\omega\text{ReTr}\left\{ \partial_{T}\tilde{\Psi}_{(0)}^{<}\partial_{\omega}\tilde{G}_{(0)}^{A}+\partial_{T}\tilde{\Psi}_{(0)}^{R}\partial_{\omega}\tilde{G}_{(0)}^{<} \right.
  \left.-\partial_{\omega}\tilde{\Psi}_{(0)}^{<}\partial_{T}\tilde{G}_{(0)}^{A}-\partial_{\omega}\tilde{\Psi}_{(0)}^{R}\partial_{T}\tilde{G}_{(0)}^{<}\right\},
  \label{first_order}
\end{multline}
where $\text{ReTr}$ denotes the real part of the trace. We find that $\tau_{(1)}$ is proportional to $\dot{\theta}$ and as a result, it can be alternately expressed as
\begin{equation}
\tau_{(1)} = -\xi (\theta) \dot{\theta},
\end{equation}
where $\xi$ is the electronic viscosity coefficient. Thus, (\ref{first_order}) denotes the dissipative frictional torque.

The  fluctuations about the average torque are treated as a Gaussian stochastic variable which is quantified entirely by its first two moments:
\begin{equation}
\langle \delta \hat\tau(t) \rangle = 0, \;\;\;\;\;\;\; \langle \delta \hat\tau(t) \delta \hat\tau(t') \rangle = D \delta (t-t'),
\label{Dcorr}
\end{equation}
where $D$ is the electronic diffusion coefficient which we aim to find an expression for. Note that we have taken the white-noise approximation such that the stochastic force is delta-correlated. 

Here, we provide a final expression for $D$, while the derivation follows  Ref. \cite{preston2020}:

\begin{align}
D(\theta) & =\frac{1}{2\pi}\int d\omega \text{Tr}\left\{\tilde{G}_{(0)}^{>}\tilde{\Phi}_{(0)}^{A}\tilde{G}_{(0)}^{<}\tilde{\Phi}_{(0)}^{A}+\tilde{G}_{(0)}^{R}\tilde{\Phi}_{(0)}^{>}\tilde{G}_{(0)}^{<}\tilde{\Phi}_{(0)}^{A}\right.
  +\tilde{G}_{(0)}^{>}\tilde{\Phi}_{(0)}^{A}\tilde{G}_{(0)}^{R}\tilde{\Phi}_{(0)}^{<}+\tilde{G}_{(0)}^{R}\tilde{\Phi}_{(0)}^{>}\tilde{G}_{(0)}^{R}\tilde{\Phi}_{(0)}^{<}+\tilde{\Psi}_{(0)}^{>}\tilde{G}_{(0)}^{A}\tilde{\Psi}_{(0)}^{<}\tilde{G}_{(0)}^{A}
  \nonumber\\
 &+\tilde{\Psi}_{(0)}^{R}\tilde{G}_{(0)}^{>}\tilde{\Psi}_{(0)}^{<}\tilde{G}_{(0)}^{A}
 \left.+\tilde{\Psi}_{(0)}^{>}\tilde{G}_{(0)}^{A}\tilde{\Psi}_{(0)}^{R}\tilde{G}_{(0)}^{<}+\tilde{\Psi}_{(0)}^{R}\tilde{G}_{(0)}^{>}\tilde{\Psi}_{(0)}^{R}\tilde{G}_{(0)}^{<}+\tilde{G}_{(0)}^{>}\tilde{\zeta}_{(0)}^{<}+\tilde{\zeta}_{(0)}^{>}\tilde{G}_{(0)}^{<}\right.
 +\tilde{G}_{(0)}^{>}\tilde{\Psi}_{(0)}^{<}\tilde{G}_{(0)}^{A}\tilde{\Phi}_{(0)}^{A}
   \nonumber\\
 &
 +\tilde{\Psi}_{(0)}^{>}\tilde{G}_{(0)}^{A}\tilde{\Phi}_{(0)}^{A}\tilde{G}_{(0)}^{<}
 +\tilde{G}_{(0)}^{>}\tilde{\Psi}_{(0)}^{R}\tilde{G}_{(0)}^{<}\tilde{\Phi}_{(0)}^{A} \left. +\tilde{\Psi}_{(0)}^{R}\tilde{G}_{(0)}^{>}\tilde{\Phi}_{(0)}^{A}\tilde{G}_{(0)}^{<}+\tilde{G}_{(0)}^{>}\tilde{\Psi}_{(0)}^{R}\tilde{G}_{(0)}^{R}\tilde{\Phi}_{(0)}^{<}+\tilde{\Psi}_{(0)}^{R}\tilde{G}_{(0)}^{R}\tilde{\Phi}_{(0)}^{>}\tilde{G}_{(0)}^{<} \right\} ,
  \label{D}
\end{align}
where we have introduced an additional self-energy-like term, defined as ($c=<,>, R, A$)
\begin{equation}
\zeta^{c}_{ij}(t,t') = \sum_{k\alpha} \partial_\theta t_{ik\alpha}(\theta(t))g_{ k\alpha}^{c} (t,t') \partial_\theta t_{k\alpha j}(\theta(t')),
\end{equation}
whose perturbative expansion is defined in the usual way. The diffusion coefficient according to (\ref{D}) then gives a means of quantifying the stochastic force in numerical simulations.

\section{Solving for the Adiabatic and First Order Green's Functions}

What remains is to calculate explicit expressions for both the adiabatic and first order Green's functions, as well as the self-energy-like terms, in the frequency domain. The Green's functions evolve according to the Keldysh-Kadanoff-Baym equations, given in the Wigner space as \cite{preston2020,preston21,preston22}
\begin{equation}
\Big(\omega+\frac{i}{2}\partial_{T}-e^{\frac{1}{2i}\lambda\partial_{\omega}^{G}\partial_{T}^{h}} h(T)\Big)\tilde{G}^{R/A}=I
+e^{\frac{1}{2i}\lambda\left(\partial_T^\Sigma \partial_\omega^G-\partial_{\omega}^{\Sigma}\partial_{T}^{G}\right)}\tilde{\Sigma}^{R/A}\tilde{G}^{R/A},\label{eqm4}
\end{equation}
\begin{equation}
\Big(\omega+\frac{i}{2}\partial_{T}-e^{\frac{1}{2i}\lambda\partial_{\omega}^{G}\partial_{T}^{h}} h(T)\Big)\tilde{G}^{</>}=e^{\frac{1}{2i}\lambda\left(\partial_T^\Sigma \partial_\omega^G-\partial_{\omega}^{\Sigma}\partial_{T}^{G}\right)}
\Big(\tilde{\Sigma}^{R}\tilde{G}^{</>}+\tilde{\Sigma}^{</>}\tilde{G}^{A}\Big),\label{eqm3}
\end{equation}
where we have shown the retarded/advanced and the lesser/greater terms collectively. Here, we adopt the convenient notation for derivatives, $\partial_T^G$, which denotes a partial derivative acting on the $G$ term with respect to $T$, and so on. We have once again introduced the book-keeping parameter, $\lambda$, for clarity in our perturbative expansions. The self-energies take the conventional form ($c=<,>, R, A$):
\begin{equation}
\Sigma^{c}_{ij}(t,t') = \sum_{k\alpha} t_{ik\alpha}(\theta(t))g_{ k\alpha}^{c} (t,t') t_{k\alpha j}(\theta(t')),
\end{equation}
and we apply our usual ansatz to the self-energies,
\begin{equation}
\tilde{\Sigma}=\tilde{\Sigma}_{(0)} + \lambda\tilde{\Sigma}_{(1)} + \lambda^2\tilde{\Sigma}_{(2)}+ ....
\label{ansatz_sigma}
\end{equation}
To solve for the form of the adiabatic and first-order Green's functions, we take a perturbative expansion of the exponentials in (\ref{eqm4}) and (\ref{eqm3}) as well as substituting in our perturbative ansatzes, (\ref{ansatz_G}) and (\ref{ansatz_sigma}). Truncating after the zeroth order and solving for $\tilde{G}_{(0)}$ yields the standard adiabatic Green's functions as follows:
\begin{equation}
\tilde{G}_{(0)}^{R/A} = \Big(\omega I-h-\tilde{\Sigma}_{(0)}^{R/A}\Big)^{-1}, 
\label{GRA}
\end{equation}
\begin{equation}
\tilde{G}_{(0)}^{</>} = \tilde{G}_{(0)}^R \tilde{\Sigma}_{(0)}^{</>} \tilde{G}_{(0)}^A.
\label{GLG}
\end{equation}
For the first-order, we consider terms linear in $\lambda$ such that we obtain
\begin{equation}
\tilde{G}_{(1)}^{R/A} = \frac{1}{2i}\tilde{G}_{(0)}^{R/A} \Big[\tilde{G}_{(0)}^{R/A}, \partial_T h \Big] \tilde{G}_{(0)}^{R/A},
\end{equation}
\begin{equation}
\tilde{G}_{(1)}^{</>} = \tilde{G}_{(0)}^R \tilde{\Sigma}_{(0)}^{</>} \tilde{G}_{(1)}^A + \tilde{G}_{(1)}^R \tilde{\Sigma}_{(0)}^{</>} \tilde{G}_{(0)}^A 
+ \frac{1}{2i} \tilde{G}_{(0)}^R \Big(\partial_{T}h \tilde{G}_{(0)}^R \partial_{\omega} \tilde{\Sigma}^{</>} + \tilde{G}_{(0)}^{</>} \partial_T h + h.c \Big)\tilde{G}^A_{(0)}.
\end{equation}
We now solve for the adiabatic and first-order components of the self-energy-like terms. Rather than considering each variant of self-energy individually, we will instead consider the following more general expression ($c=<,>, R, A$)
\begin{equation}
\Xi^{c}_{\alpha,ii'} = \sum_k A_{ik\alpha}(t) g^{c}_{k\alpha}(t,t') B_{k\alpha i'}(t'),
\label{GSE}
\end{equation}
where $A$ and $B$ are arbitrary functions of time. Obviously, when $A_{k\alpha i}=B_{k \alpha i}=t_{k\alpha i}$, we obtain $\Sigma^{c}$, while different choices allow us to obtain $\Psi$, $\Phi$ and $\zeta$. We apply the Wigner transform to the above while making use of the {   shift operator, defined according to $f(x+h)=e^{hd^f_x}f(x)$ where we use $d^f_x$ to denote the derivative with respect to $x$ which acts on $f$ (to avoid ambiguity) , to obtain}
\begin{align}
\tilde{\Xi}^c_{\alpha,ii'} 
&= \sum_k \int^\infty_{-\infty} d\tau e^{i\omega \tau} e^{\frac{\tau}{2}d^A_T}A_{i k\alpha}(T) g^c_{k\alpha}(t,t')e^{\frac{-\tau}{2}d^B_T}B_{k\alpha i'}(T)\\
&= \sum_k \int^\infty_{-\infty} d\tau e^{i\omega \tau} e^{\frac{1}{2i}\overleftarrow{\partial_\omega^e}(d^A_T - d^B_T)}A_{ik\alpha}(T) 
g^c_{k\alpha }(t,t')B_{k\alpha i'}(T),
\end{align}
where the $\overleftarrow{\partial_\omega^e}$ notation denotes the derivative operator acting to the left on the exponential. Now we take all the terms that are independent of $\tau$ outside of the integral, leaving us with
\begin{align}
\tilde{\Xi}^c_{\alpha,ii'} &=\sum_k e^{\frac{1}{2i}\overrightarrow{\partial_\omega^e}(d^A_T - d^B_T)}A_{i k\alpha}(T)B_{k\alpha i'}(T) \int^\infty_{-\infty} d\tau e^{i\omega \tau}  g^c_{k\alpha}(t,t') 
\\&=\sum_k e^{\frac{1}{2i}\overrightarrow{\partial_\omega^G}(d^A_T - d^B_T)}A_{i k\alpha}(T)B_{k\alpha i'}(T)  \tilde{g}^c_{ k\alpha}(T,\omega).
\end{align}
Finally, we take a power series expansion of the exponential to find
\begin{align}
\tilde{\Xi}^c_{\alpha,ii'} &= \sum_k A_{ik\alpha} \tilde{g}^c_{ k\alpha} B_{k\alpha i'}
+ \frac{1}{2i}\sum_k \frac{\partial\tilde{g}^c_{ k\alpha}}{\partial \omega} \left( \frac{dA_{ik\alpha}}{dT} B_{k\alpha i'} - A_{ik\alpha} \frac{dB_{k\alpha i'}}{dT}\right) + ...= \tilde{\Xi}^c_{(0),\alpha,ii'} + \tilde{\Xi}^c_{(1),\alpha,ii'} + ...,
\label{GSE_pert}
\end{align}
where the functional dependencies are clear from the context. Thus, (\ref{GSE_pert}) allows us to calculate each  of the required orders of self-energy-like terms. 
If we consider $A_{k\alpha i}=B_{k \alpha i}=t_{k\alpha i}$, the adiabatic component corresponds to the standard self-energy. 
We make the wide-band approximation for the electrodes. The retarded/advanced component is given by
\begin{equation}
\tilde{\Sigma}_{(0),\alpha,ii'}^{R/A}=\mp\frac{i}{2}\Gamma_{\alpha,ii'},
\end{equation}
where the level-broadening takes the form
\begin{equation}
\Gamma_{\alpha,ii'} =2\pi t_{\alpha i}^{*}t_{\alpha i'}\rho_{\alpha},
\end{equation}
where density of states $\rho$ is a constant and $t_{\alpha k i} =t_{\alpha i} $ under the wide-band approximation. 
The equation for the lesser case takes the form
\begin{equation}
\tilde{\Sigma}_{(0),\alpha,ii'}^{<}(\omega,T)=if_\alpha(\omega)\Gamma_{\alpha,ii'}(T),
\end{equation}
where $f_\alpha(\omega)$ is the Fermi-Dirac distribution;
\begin{equation}
f_\alpha (\omega) = \frac{1}{e^{\frac{\omega - \mu_\alpha}{k_B T_\alpha}} + 1}.
\end{equation}
Here, $\mu_\alpha$ is the chemical potential for the $\alpha$ lead while $T_\alpha$ is the macroscopic temperature and $k_B$ is Boltzmann's constant. The form of $\Psi$, $\Phi$ and $\zeta$ can be found equivalently by replacing $\Gamma$ in the above equations with, $\Gamma^\Psi$, $\Gamma^\Phi$ and $\Gamma^\zeta$, respectively, as given by
\begin{equation}
\Gamma^\Psi_{\alpha,ii'} =2\pi \partial_\theta t_{\alpha i}^{*}t_{\alpha i'}\rho_{\alpha},
\end{equation}
\begin{equation}
\Gamma^\Phi_{\alpha,ii'} =2\pi t_{\alpha i}^{*} \partial_\theta t_{\alpha i'}\rho_{\alpha},
\end{equation}
\begin{equation}
\Gamma^\zeta_{\alpha,ii'} =2\pi \partial_\theta t_{\alpha i}^{*}\partial_\theta t_{\alpha i'}\rho_{\alpha}.
\end{equation}
Under the wide-band approximation, $\tilde{\Xi}_{(1)}^{R/A} = 0$, and so we need only consider the lesser case.

\end{widetext}

\clearpage

\end{document}